%% file: main.tex
\documentclass[conference]{IEEEtran}
\IEEEoverridecommandlockouts
\usepackage{cite}
\usepackage{amsmath,amssymb,amsfonts}

\usepackage{graphicx}
\usepackage{textcomp}
\usepackage{xcolor}
\def\BibTeX{{\rm B\kern-.05em{\sc i\kern-.025em b}\kern-.08em
    T\kern-.1667em\lower.7ex\hbox{E}\kern-.125emX}}

\usepackage{pdfpages}
\usepackage{epsfig}
\usepackage{color}
\usepackage{amsxtra}
\usepackage{enumitem}
\usepackage{acro}
\usepackage[T1]{fontenc}
\usepackage{setspace}

\usepackage[cmintegrals]{newtxmath}
\usepackage{graphicx}
\graphicspath{{Figures/}}
\usepackage{url}
\usepackage{cite}
\usepackage{algorithm}
\usepackage{algpseudocode}

\makeatletter
\def\BState{\State\hskip-\ALG@thistlm}
\makeatother

\usepackage{epstopdf}

\usepackage{amsxtra}
\usepackage{amstext}

\usepackage{latexsym}
\usepackage{dsfont} 
\usepackage{color}
\usepackage{multirow}
\usepackage{float}
\usepackage[bookmarks=false]{hyperref}
\usepackage{eucal}
\usepackage{tabularx,colortbl}

\usepackage{lscape}

\usepackage{algorithm}
\usepackage{algpseudocode}

\usepackage{units}
\usepackage{cases}
\usepackage[ subrefformat=parens,labelformat=parens, caption=false, font=footnotesize]{subfig}
\usepackage{mathtools}
\usepackage{pgfplots}
\usepackage{pgf}
\usepackage{tikz}
\pgfplotsset{compat=1.15}

\usepackage{mathtools}

\DeclareAcronym{ris}{
  short = RIS,
  long  = reconfigurable intelligent surface
}

\DeclareAcronym{mmse}{
  short = MMSE,
  long = minimum mean square error
}
\DeclareAcronym{ls}{
  short = LS,
  long = least square
}

\DeclareAcronym{RS-LS}{
  short = RS-LS,
  long = reduced-subspace least squares
}

\DeclareAcronym{mimo}{
  short = MIMO,
  long =  multiple-input multiple-output
}
\DeclareAcronym{BS}{
  short = BS,
  long =  base station
}
\DeclareAcronym{UE}{
  short = UE,
  long =  user equipment
}

\DeclareAcronym{MSE}{
  short = MSE,
  long = mean square error  
}

\DeclareAcronym{TDD}{
  short = TDD,
  long = time-division duplex  
}

\DeclareAcronym{PDF}{
  short = PDF,
  long = probability density function  
}

\DeclareAcronym{ML}{
  short = ML,
  long = maximum likelihood 
}

\DeclareAcronym{UPA}{
  short = UPA,
  long = uniform planner array 
}

\DeclareAcronym{los}{
  short = LoS,
  long = line-of-sight 
}

\DeclareAcronym{MC}{
  short = MC,
  long = mutual coupling 
}
\DeclareAcronym{NMSE}{
  short = NMSE,
  long = normalized mean squared error
}

\DeclareAcronym{iid}{
  short = IID,
  long = independent and identically distributed
}

\DeclareAcronym{snr}{
  short = SNR,
  long = signal-to-noise ratio
}

\makeatother

\makeatletter
\def\ps@IEEEtitlepagestyle{%
  \def\@oddfoot{\mycopyrightnotice}%
  \def\@oddhead{\hbox{}\@IEEEheaderstyle\leftmark\hfil\thepage}\relax
  \def\@evenhead{\@IEEEheaderstyle\thepage\hfil\leftmark\hbox{}}\relax
  \def\@evenfoot{}%
}

\def\mycopyrightnotice{%
  \begin{minipage}{\textwidth}
  \centering \scriptsize
    This work has been accepted for presentation at the IEEE GLOBECOM Conference. Copyright may be transferred without notice, after which this version may no longer be accessible.
  \end{minipage}
}

\begin{document}
\include{macros} 
\bstctlcite{IEEEexample:BSTcontrol}

\title{{RIS-Aided Near-Field Channel Estimation under Mutual Coupling and Spatial Correlation
}
{\footnotesize}
\thanks{This work is supported by the AUB's University Research Board and Vertically Integrated Projects Program, and KAUST's Office of Sponsored Research under Award No. ORFS-CRG12-2024-6478.}
}
\author{Ahmad~Dkhan$^{\dagger}$, Simon~Tarboush$^{*}$, Hadi~Sarieddeen$^{\dagger}$, Tareq Y. Al-Naffouri$^{*}$ \\
\small$^{\dagger}$ Department of Electrical and Computer Engineering, American University of Beirut, Beirut, Lebanon,\\
$^{*}$ Department of CEMSE, Information Science Lab, King Abdullah University of Science and Technology, Thuwal, Saudi Arabia\\
amd53@mail.aub.edu, simon.w.tarboush@gmail.com, hs139@aub.edu.lb, tareq.alnaffouri@kaust.edu.sa
}

\maketitle
\begin{abstract}
The integration of reconfigurable intelligent surfaces (RIS) with extremely large multiple-input multiple-output (MIMO) arrays at the base station has emerged as a key enabler for enhancing wireless network performance. However, this setup introduces high-dimensional channel matrices, leading to increased computational complexity and pilot overhead in channel estimation. Mutual coupling (MC) effects among densely packed unit cells, spatial correlation, and near-field propagation conditions further complicate the estimation process. Conventional estimators, such as linear minimum mean square error (MMSE), require channel statistics that are challenging to acquire for high-dimensional arrays, while least squares (LS) estimators suffer from performance limitations. To address these challenges, the reduced-subspace least squares (RS-LS) estimator leverages array geometry to enhance estimation accuracy. This work advances the promising RS-LS estimation algorithm by explicitly incorporating MC effects into the more realistic and challenging near-field propagation environment within the increasingly relevant generalized RIS-aided MIMO framework. Additionally, we investigate the impact of MC on the spatial degrees of freedom (DoF). Our analysis reveals that accounting for MC effects provides a significant performance gain of approximately 5 dB at an SNR of 5 dB, compared to conventional methods that ignore MC.
\end{abstract}


\section{Introduction}

\Acp{ris} have gained significant attention for their ability to dynamically shape wireless propagation by adjusting the impedance of unit cells to control signal reflections and transmissions \cite{Basar2019Wireless, Wu2019Intelligent, Di2020Smart}. Such signal shaping enhances coverage, signal reliability, and overall communication performance. Accurate channel modeling is crucial, particularly for large-scale \acp{ris}, where the expanded surface aperture extends the near-field region \cite{Cui2023Near}. Unlike far-field conditions, near-field channels exhibit spherical wave propagation and spatial non-stationarity \cite{Cui2023Near,Tarboush2023Compressive}. 

Additionally, the \ac{MC} effect in dense \ac{mimo} arrays and \acp{ris} introduces electromagnetic interactions that distort radiation patterns, alter channel characteristics, and intensify spatial correlation, leading to performance bottlenecks \cite{Janaswamy2002Effect, Chen2018Review}. Studies such as \cite{Sun_2022} investigate \ac{MC} effects on holographic \acp{ris}, analyzing their influence on eigenvalues, spatial correlation, and array gain. Moreover, neglecting \ac{MC} leads to significant estimation errors and system degradation, as demonstrated in recent works~\cite{KolomvakisExploiting,Zheng2024On,Zheng2024Mutual}. The authors in \cite{KolomvakisExploiting} highlight that \ac{MC} critically affects \ac{mimo} channel estimation, even more than spatial correlation. Similarly, \cite{Zheng2024On} demonstrates that in active \acp{ris} with large amplitudes, \ac{MC} can significantly degrade estimation performance in tightly spaced or large-scale setups. The study in \cite{Zheng2024Mutual} highlights the crucial role of accounting for \ac{MC} in \ac{ris}-aided channel estimation and beamforming. Despite extensive studies on \ac{MC}, its impact in the near-field remains largely unexplored for large arrays and surfaces.

Unlocking the potential of \ac{ris} technology depends on accurate channel estimation, which is challenging due to the unique characteristics of \ac{ris}-aided links \cite{Pan2022Overview}. Deploying numerous reflecting surfaces alongside \ac{mimo} arrays at the base station results in high-dimensional channel matrices, increasing estimation overhead and computational complexity~\cite{Jensen2020Optimal}. The \ac{mmse} estimator requires perfect knowledge of channel statistics, which is difficult to obtain in high-dimensional settings, while \ac{ls} estimation, though independent of prior information, performs poorly, especially at low \ac{snr}~\cite{Jensen2020Optimal}. To address these challenges, \cite{Demir2022Exploiting} proposed the \ac{RS-LS} estimator, which leverages channel geometry to reduce dimensionality and improve estimation accuracy in far-field \ac{ris}-aided systems; its effectiveness in near-field \ac{mimo} systems was later explored in \cite{Demir2024Spatial}. 

The potential of the \ac{RS-LS} estimator to exploit \ac{MC} information in a generalized MIMO-RIS setting with accurate near-field modeling is yet to be investigated. This paper addresses this gap by integrating \ac{MC} effects and near-field characteristics into the channel estimation process. We analyze their combined impact on linear estimators and propose an enhanced \ac{RS-LS} estimator that leverages \ac{MC} information to improve estimation accuracy. We further derive the general error covariance matrix for \ac{ris}-aided \ac{mimo} systems and examine the degradation of spatial degrees of freedom (DoF) due to strong \ac{MC} in dense antenna arrays, highlighting its implications for estimation performance. Our analysis shows that integrating \ac{MC} into the \ac{RS-LS} estimator improves accuracy by approximately $\unit[5]{dB}$ at an \ac{snr} of $\unit[5]{dB}$ compared to the \ac{RS-LS} estimator without \ac{MC} consideration. Furthermore, we identify the most promising scenarios for the \ac{RS-LS} estimator, demonstrating that its highest performance gain occurs at small inter-element spacing.

Regarding notation, scalars $(a, A)$, vectors $(\av)$, and matrices $(\Am)$ are represented by non-bold, bold lowercase, and bold uppercase letters, respectively. $\mbf{I}_M$ is an identity matrix of size $M\times M$. $(\cdot)^\Tpow$, $(\cdot)^\Strpow$, $(\cdot)^\Hpow$, and ${(\cdot)}^{-1}$ stand for the transpose, conjugate, Hermitian, and inverse operators, respectively. $\mathbf{A} \odot \mathbf{B}$, $\mathbf{A} \otimes \mathbf{B}$, and $\mathbf{A} \diamond \mathbf{B}$ denote the Hadamard, Kronecker, and Khatri–Rao products, respectively. $\EE[\cdot]$ denotes the expectation operator. $\diagopr{a_1,a_2,\dots,a_N}$ is an $N\times N$ diagonal matrix with diagonal entries $\{a_1,a_2,\dots,a_N\}$. The floor and ceiling operations are denoted by $\lfloor \cdot \rfloor$ and $\lceil \cdot \rceil$, respectively. The modulo operation, \(\mathrm{mod}(a, b)\), returns the remainder of the division of $a$ by 
$b$.

\section{System Model}

We analyze the uplink of a \ac{TDD} \ac{ris}-aided \ac{mimo} communication system, where a \ac{UE} with \( N = N_\Ht \times N_\Vt \) antennas communicates with a \ac{BS} with \( M = M_\Ht \times M_\Vt \) antennas through a nearly-passive \ac{ris} consisting of \( K = K_\Ht \times K_\Vt \) unit cells. We assume the direct \ac{UE}-\ac{BS} link is blocked, focusing entirely on the role of the \ac{ris} in signal propagation and channel estimation.

\begin{figure}[t]
   \centering
   \includegraphics[width=0.95\linewidth]{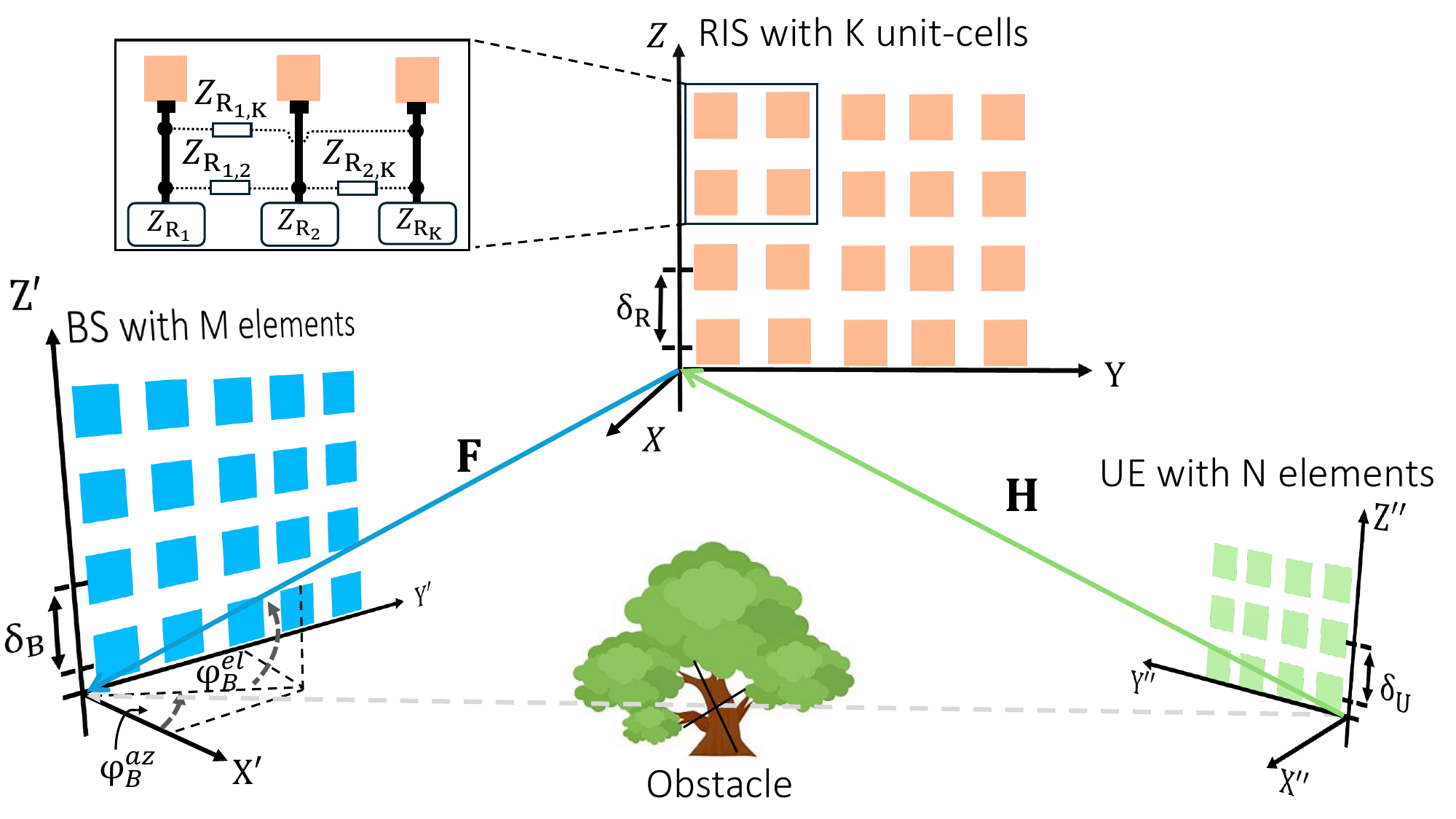}
    \caption{An \ac{ris}-aided uplink MIMO communication system model.}
    \label{fig:sys}
\end{figure}

We assume a Cartesian coordinate system where the \ac{ris}, \ac{BS}, and \ac{UE} elements form a \ac{UPA} on the Y-Z plane, as shown in Fig.~\ref{fig:sys}. The position of the $k$th unit cell of the \ac{ris} ($k \!\in\! \{1, \dots, K\}$), the $m$th antenna of the \ac{BS} ($m \!\in\! \{1, \dots, M\}$), and the $n$th antenna of the \ac{UE} ($n \!\in\! \{1, \dots, N\}$) are given by $\pv_k = [ 0 , i_k \delta_{\Rt} , j_k \delta_{\Rt} ]$, $\pv_m = [ 0 , i_m \delta_{\Bt} , j_m \delta_{\Bt} ]$, and $\pv_n = [ 0 , i_n \delta_{\Ut} , j_n \delta_{\Ut} ]$, where $i_k = \text{mod}(k \!-\! 1, K_\Ht)$, $j_k = \left\lfloor (k \!-\! 1) / {K_\Ht} \right\rfloor$, $i_m = \text{mod}(m \!-\! 1, M_\Ht)$, $j_m = \left\lfloor (m \!-\! 1) / {M_\Ht} \right\rfloor$, $i_n = \text{mod}(n \!-\! 1, N_\Ht)$, and $j_n = \left\lfloor (n \!-\! 1) / {N_\Ht} \right\rfloor$. Here, $\delta_{\Rt}$, $\delta_{\Bt}$, and $\delta_{\Ut}$ represent the inter-element spacings for the \ac{ris}, \ac{BS}, and \ac{UE}, respectively, and $K_\Ht$, $M_\Ht$, and $N_\Ht$ are the number of elements along the horizontal axis for each.

We model a narrowband channel with block fading, assuming static channels within coherence intervals. During uplink channel training, the phase shifts of the \ac{ris} elements are adjusted. At the $\ell$th training step ($\ell \in \{1, \dots, T_\ppow\}$), with training duration, $T_\ppow$, shorter than the channel coherence time, $T_{\cpow\hpow}$, the \ac{BS} received signal, $\yv_\ell \!\in\! \mathbb{C}^M$, is given by \cite{Swindlehurst2022Channel, Pan2022Overview}
\begin{equation}
\label{MRIs}
\yv_\ell = \Fm \diagopr{\phiv_\ell} \Hm \xv_\ell + \nv_\ell = (\xv_\ell^\Tpow \otimes \mbf{I}_M )(\Hm^\Tpow \diamond \Fm) \phiv_\ell + \nv_\ell,
\end{equation}
where $\xv_\ell \!\in\! \mathbb{C}^N$ represents the \ac{UE} transmitted pilot symbols, $\nv_\ell \sim \mathcal{CN}(\mathbf{0}, \sigma^2 \mathbf{I}_M)$ represents the noise, which follows a circularly symmetric complex Gaussian random variables, and $\Fm \in \mathbb{C}^{M \times K}$ and $\Hm \in \mathbb{C}^{K \times N}$ denote the \ac{ris}-\ac{BS} and \ac{UE}-\ac{ris} channels, respectively (see Sec.~\ref{sec:SP} for details). The \ac{ris} phase shift vector is \(\phiv_\ell = [\phi_{\ell,1}, \ldots, \phi_{\ell,K}]^\Tpow = [e^{j\nu_{\ell,1}}, \ldots, e^{j\nu_{\ell,K}}]^\Tpow \in \mathbb{C}^K\), where $\nu_{\ell,k} \in [0, 2\pi)$. We express a cascaded channel as \(\Cm = [\cv_1, \cv_2, \dots, \cv_K] \triangleq \Hm^\Tpow \diamond \Fm \in \mathbb{C}^{N M \times K}\). Using the Kronecker product property ($\text{vec}(ABC) \!=\! (C^T \!\otimes\! A) \text{vec}(B)$), and defining $\cv\!\triangleq\!\text{vec}(\Cm)$, \eqref{MRIs} is expressed as \cite{Swindlehurst2022Channel}
\begin{equation}
\yv_\ell = \left(\phiv_\ell^\Tpow \otimes \xv_\ell^\Tpow \otimes \mbf{I}_M \right) \cv + \nv_\ell = \Qm_\ell \cv + \nv_\ell.
\end{equation}
Combining received data from $T_\ppow$ training steps results in
\begin{equation}
\yv = \Qm \cv + \nv,\label{meq}
\end{equation}
\begin{equation}
\begin{aligned}
\yv &= \begin{bmatrix} \yv_1 \\ \yv_2 \\ \vdots \\ \yv_{T_\ppow} \end{bmatrix}, \quad
\Qm = \begin{bmatrix} \Qm_1 \\ \Qm_2 \\ \vdots \\ \Qm_{T_\ppow} \end{bmatrix}, \quad
\cv = \begin{bmatrix} \cv_1 \\ \cv_2 \\ \vdots \\ \cv_K \end{bmatrix}, \quad
\nv = \begin{bmatrix} \nv_1 \\ \nv_2 \\ \vdots \\ \nv_{T_\ppow} \end{bmatrix},
\end{aligned}
\end{equation}
where $\yv$, $\Qm$, $\cv$, and $\nv$ represent the observation, training matrix, unknown channel, and noise vector, respectively. To ensure that \( \Qm \) is full rank, enabling accurate estimation of \( \cv \), the training duration must satisfy \(T_\ppow \geq K N.\)

\label{sec:sys_mod}

A common training approach divides the total period $T_\ppow$ into $T_\ppow / N$ intervals \cite{Swindlehurst2022Channel}, each of length $N$. For simplicity, $T_\ppow / N$ is assumed to be an integer. Within each interval, $\phiv_\ell$ remains constant, and the phase shift matrix is organized as follows:

\begin{equation}
\Phim =
\begin{bmatrix}
\phi_{1,1} & \phi_{1,2} & \dots & \phi_{1,K} \\ 
\vdots & \vdots & \ddots & \vdots \\ 
\phi_{1,1} & \phi_{1,2} & \dots & \phi_{1,K} \\ 
\vdots & \vdots & \ddots & \vdots \\ 
\phi_{T_\ppow/N,1} & \phi_{T_\ppow/N,2} & \dots & \phi_{T_\ppow/N,K} \\ 
\vdots & \vdots & \ddots & \vdots \\ 
\phi_{T_\ppow/N,1} & \phi_{T_\ppow/N,2} & \dots & \phi_{T_\ppow/N,K}
\end{bmatrix}
\!\raisebox{30pt}{%
$\left. \rule{0pt}{20pt} \right\} 
\shortstack{\text{\small $\phiv_1$ is repeated $N$ times} \\ \text{\small in the first interval}}$%
}.
\end{equation} 

The pilot signals $\xv_\ell$ form an orthonormal sequence within each interval \cite{Swindlehurst2022Channel}. The orthonormal pilot matrix for each interval, \( \Xm \!=\! [\xv_1 \dots \xv_N] \in  \mathbb{C}^{N \times N}\), is repeated $T_\ppow / N$ times during the training period to maintain effective channel estimation and orthogonality between intervals \cite{Swindlehurst2022Channel}.

\begin{figure*}[htb] 
    \centering
    \begin{equation}
    \begin{aligned}
    \Rm_{\Ht}= \EE\left[\hv\hv^\Hpow\right]=\beta \int_{d_{\Ut}}  
    \int_{\Omega_{\Ut}} \int_{d_{\Rt}} 
    \int_{\Omega_{\Rt}} 
    f(\varphiv_{\Rt}, d_{\Rt}, \varphiv_{\Ut}, d_{\Ut}\!) \,\av_\Rt(\varphiv_{\Rt},d_{\Rt}\!)\av_\Rt^\Hpow(\varphiv_{\Rt},d_{\Rt}\!)\otimes\av_\Ut(\varphiv_{\Ut},\!d_{\Ut})\av_\Ut^\Hpow(\varphiv_{\Ut},\!d_{\Ut})\,\partial \varphiv_{\Rt} \,\partial d_{\Rt} \,\partial \varphiv_{\Ut} \,\partial d_{\Ut}\,
    \end{aligned}
    \label{corr}
    \end{equation}
\hrulefill
\vspace*{0pt}
\end{figure*}

\section{Proposed channel modeling}\label{sec:SP}

Unlike conventional far-field \ac{ris}-aided channel models \cite{Demir2022Channel,Demir2022Exploiting}, we present a comprehensive channel model incorporating a spatial correlation matrix that accounts for near-field and mutual coupling effects.

\subsection{Near-Field Channel Modeling}
Extending the model in \cite{Sayeed2002Deconstructing,Demir2024Spatial}, we define the \ac{UE}-\ac{ris} \ac{mimo} near-field channel as
\begin{equation}
\begin{split}
\Hm & =  \int_{d_{\Ut}}  
\int_{\Omega_{\Ut}} \int_{d_{\Rt}}  
\int_{\Omega_{\Rt}} 
    g(\varphiv_{\Rt}, d_{\Rt}, \varphiv_{\Ut}, d_{\Ut}) \\ 
     & \quad \av_{\Rt}(\varphiv_{\Rt}, d_{\Rt}) \av_{\Ut}(\varphiv_{\Ut}, d_{\Ut})
     \, \partial \varphiv_{\Rt}  \,
     \partial d_{\Rt} \, 
     \partial \varphiv_{\Ut} \,
     \partial d_{\Ut},
\end{split} \label{H}
\end{equation}
where \( g(\varphiv_{\Rt}, d_{\Rt}, \varphiv_{\Ut}, d_{\Ut}) \) is the angular and distance spreading function that defines the gain and phase shift induced by a scatterer at location \( (\varphiv_{\Ut}, d_{\Ut}) \) or \( (\varphiv_{\Rt}, d_{\Rt}) \) as observed from the \ac{UE} or \ac{ris}, respectively. Here, \( \varphiv \!=\! [\varphi^{\apow\zpow}, \varphi^{\epow\lpow}]^\Tpow\) represents the angular coordinates, where \( \varphi^{\apow\zpow} \) and \( \varphi^{\epow\lpow} \) are the azimuth and elevation angles, respectively; the set of angles is \(\Omega \!=\! \left\{ (\varphi^{\apow\zpow}, \varphi^{\epow\lpow}) \mid \varphi^{\apow\zpow} \in \left[ -\frac{\pi}{2}, \frac{\pi}{2} \right], \varphi^{\epow\lpow} \in \left[ -\frac{\pi}{2}, \frac{\pi}{2} \right] \right\}\). Further, \( \av_{\Rt}(\cdot) \) and \( \av_{\Ut}(\cdot) \) represent the \ac{ris} and \ac{UE} near-field array response vectors, respectively. Following \cite{Sayeed2002Deconstructing,Demir2024Spatial}, we model \( g(\varphiv_{\Rt}, d_{\Rt}, \varphiv_{\Ut}, d_{\Ut}) \) as a spatially uncorrelated circularly symmetric Gaussian stochastic process with cross-correlation
\begin{equation}
\begin{split}
&\mathbb{E} \left[ g(\varphiv_{\Rt}, d_{\Rt}, \varphiv_{\Ut}, d_{\Ut}) g^\ast(\varphiv'_{\Rt}, d'_{\Rt}, \varphiv'_{\Ut}, d'_{\Ut}) \right] =  f(\varphiv_{\Rt}, d_{\Rt}, \varphiv_{\Ut}, d_{\Ut})\\
&\delta(\varphiv_{\Rt} - \varphiv'_{\Rt}) \delta(d_{\Rt} - d'_{\Rt}) \delta(\varphiv_{\Ut} - \varphiv'_{\Ut}) \delta(d_{\Ut} - d'_{\Ut}),
\end{split}
\end{equation}
where \( \delta(\cdot) \) is the Dirac delta function and \( f(\varphiv_{\Rt}, d_{\Rt}, \varphiv_{\Ut}, d_{\Ut}) \geq 0 \) is the near-field equivalent of the far-field spatial scattering function \cite{Sayeed2002Deconstructing}, representing the joint probability density function for the azimuth/elevation angles and distances, i.e.,
\begin{equation}
\int_{d_{\Ut}} \!\int_{\Omega_{\Ut}}\!\int_{d_{\Rt}}\!\int_{\Omega_{\Rt}} \!f(\varphiv_{\Rt}, d_{\Rt}, \varphiv_{\Ut}, d_{\Ut})\, \partial \varphiv_{\Rt} \,
     \partial d_{\Rt} \, 
     \partial \varphiv_{\Ut} \,
     \partial d_{\Ut}=1.
\end{equation}
Let the scatterer-\( k \)th \ac{ris} element separation distance be \cite{Demir2024Spatial}
\begin{align}
d_{k} &= \bigg( \big( d \cos(\varphi^{\epow\lpow}) \cos(\varphi^{\apow\zpow}) \big)^2 + \big( d \sin(\varphi^{\epow\lpow}) - j_{k} \delta_{\Rt} \big)^2 \\
& + \big( d \cos(\varphi^{\epow\lpow}) \sin(\varphi^{\apow\zpow}) - i_{k} \delta_{\Rt} \big)^2 \bigg)^{1/2},
\end{align}
where \( d \) represents the distance from the origin to the respective scatterer. The near-field \ac{ris} array response vector is expressed as
\begin{equation}
\label{eq:near-steer_vec}
\av_{\Rt}(\varphiv, d) = \left[ e^{-j\frac{2\pi}{\lambda}(d_{1} - d)}, \dots, e^{-j\frac{2\pi}{\lambda}(d_{K} - d)} \right]^\Tpow.
\end{equation}
We similarly define the \ac{UE} and \ac{BS} near-field response vectors.

\subsection{Near Field Spatial Correlation}

Let \( \mathbf{h} = \text{vec}(\mathbf{H}) \), and the near-field spatial correlation matrix be computed {as \eqref{corr}, where \( \beta \!>\! 0 \) represents the average channel gain. Assuming the receive and transmit scattering environments contribute independently to the overall angular and distance distributions, the joint distribution is given by \( f(\varphiv_{\Rt},d_{\Rt},\varphiv_{\Ut},d_{\Ut}) \!=\! f_{\Rt}(\varphiv_{\Rt},d_{\Rt}) \, f_{\Ut}(\varphiv_{\Ut},d_{\Ut}) \). Thus, from \eqref{corr} and leveraging the Kronecker-based correlated channel model, the correlation matrix for channel \( \mathbf{H} \) is given by 
\begin{equation}
\Rm_\Ht=\Rm_{\Ht\Rt}\otimes\Rm_{\Ht\Ut}, 
\end{equation}
where $\Rm_{\Ht\Rt}$ and $\Rm_{\Ht\Ut}$ are the  \ac{ris} and \ac{UE} spatial correlation matrices, respectively. Similarly, the correlation matrix for the \ac{ris}-\ac{BS} channel, \(\Fm\), is
\begin{equation}
\Rm_\Ft=\Rm_{\Ft\Bt}\otimes\Rm_{\Ft\Rt}.
\end{equation}
Finally, we model the channels as complex Gaussian random variable \( =\mathbf{h} \sim \mathcal{CN}(\zerov, \Rm_{\Ht\Rt} \otimes \Rm_{\Ht\Ut}), \mathbf{f} \sim \mathcal{CN}(\zerov, \Rm_{\Ft\Bt} \otimes \Rm_{\Ft\Rt}),\) with zero mean and covariance \(\big(\Rm_{\Ht\Rt}\otimes\Rm_{\Ht\Ut}\big), \big(\Rm_{\Ft\Bt}\otimes\Rm_{\Ft\Rt}\big)\).

\subsection{Mutual Coupling}\label{sec:MC}

Accurately modeling \ac{MC} is crucial, as it directly impacts the overall channel characteristics and modifies the spatial correlation structure in \eqref{corr}, yielding the effective spatial correlation matrix, \( \Rm^{\Mt\Ct} \), that captures near-field scattering and mutual element interactions. \ac{MC} is mainly determined by the mutual impedance \( \Zm \) between array elements and is computed as \cite{D’Amico2024Holographic}
\begin{equation}
\mathbf{M} = (\mathbf{Z} + r_d \mathbf{I})^{-1},
\end{equation}
where \( r_d \!>\! 0 \) is the dissipation resistance of each antenna element. To account for the \ac{MC} effect, we express the effective channel, based on findings in ~\cite[eq. (105), (107)]{Ivrlač2010Toward}, as
\begin{equation}
\Hm^{\Mt\Ct} = \mathbf{M}_{\Ht\Rt}^{1/2} \Hm \mathbf{M}_{\Ht\Ut}^{1/2},
\end{equation}
where \( \mathbf{M}_{\Ht\Rt} \) and \( \mathbf{M}_{\Ht\Ut} \) are the transmit and receive mutual coupling matrices, respectively. This formulation demonstrates that \ac{MC} directly affects the spatial correlation properties. Thus, for \( \Hm^{\Mt\Ct} \), the corresponding correlation matrix is 
\begin{equation}
\Rm^{\Mt\Ct}_\Ht= \Rm^{\Mt\Ct}_{\Ht\Rt}\!\otimes \Rm^{\Mt\Ct}_{\Ht\Ut} = \left(\!\mathbf{M}_{\Ht\Rt}^{\frac{1}{2}} \Rm_{\Ht\Rt} \mathbf{M}_{\Ht\Rt}^{\frac{1}{2}}\!\right)\otimes \left(\!\mathbf{M}_{\Ht\Ut}^{\frac{1}{2}} \Rm_{\Ht\Ut} \mathbf{M}_{\Ht\Ut}^{\frac{1}{2}}\!\right)\!.
\end{equation}
Similarly, the correlation matrix for the channel \(\Fm\) is given by
\begin{equation}
\Rm^{\Mt\Ct}_\Ft = \Rm^{\Mt\Ct}_{\Ft\Bt}\!\otimes \Rm^{\Mt\Ct}_{\Ft\Rt} = \left(\!\mathbf{M}_{\Ft\Bt}^{\frac{1}{2}} \Rm_{\text{\Ft\Bt}} \mathbf{M}_{\Ft\Bt}^{\frac{1}{2}} \!\right)\!\otimes \left(\! \mathbf{M}_{\Ft\Rt}^{\frac{1}{2}} \Rm_{\text{\Ft\Rt}} \mathbf{M}_{\Ft\Rt}^{\frac{1}{2}}\!\right).
\end{equation}
Each element of \( \Rm^{\Mt\Ct} \) is influenced by the elements of \( \Rm \), with the coupling matrix \( \mathbf{M} \) determining the weighting. For a half-wavelength dipole, the dissipation resistance is given in~\cite[Example 2.13]{Balanis2016Antenna}. Closed-form expressions for mutual impedance are available for various dipole configurations: co-linear~\cite[eq. (8.72a-b)]{Balanis2016Antenna}, parallel-in-echelon~\cite[eq. (8.73a-b)]{Balanis2016Antenna}, and side-by-side~\cite[eq. (8.69), eq. (8.71a-b)]{Balanis2016Antenna}.

\section{Channel Estimation}
\label{sec:ch_est}

\begin{figure*}[ht]
 \centering
 \subfloat[\ac{ris} inter-element spacing \(\delta_{\Rt}=0.5\).]{\label{fig:eig_0.5} \includegraphics[width=0.48\linewidth]{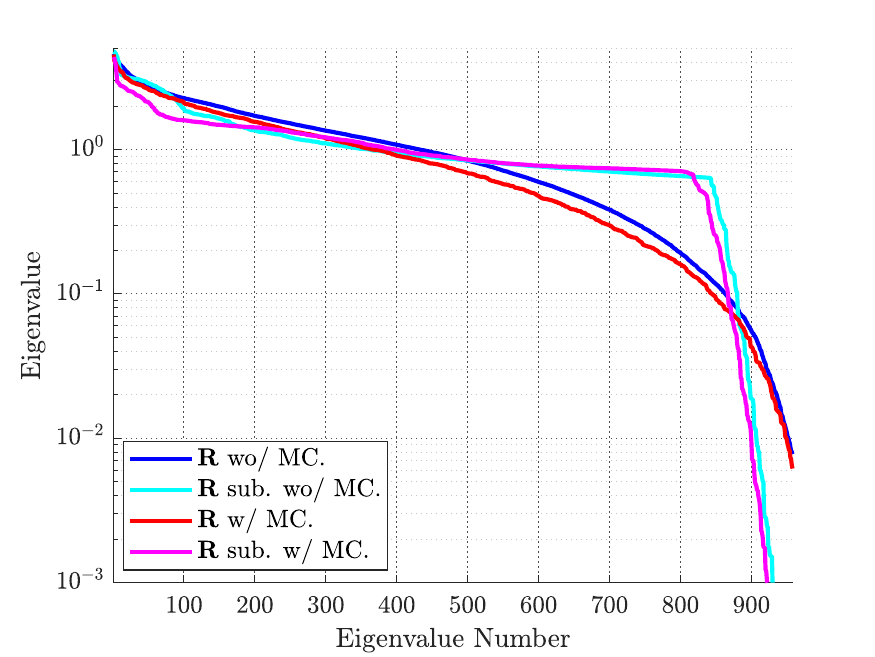}}%
 \hfill
 \subfloat[\ac{ris} inter-element spacing \(\delta_{\Rt}=0.25\).]{\label{fig:eig_0.25} \includegraphics[width=0.48\linewidth]{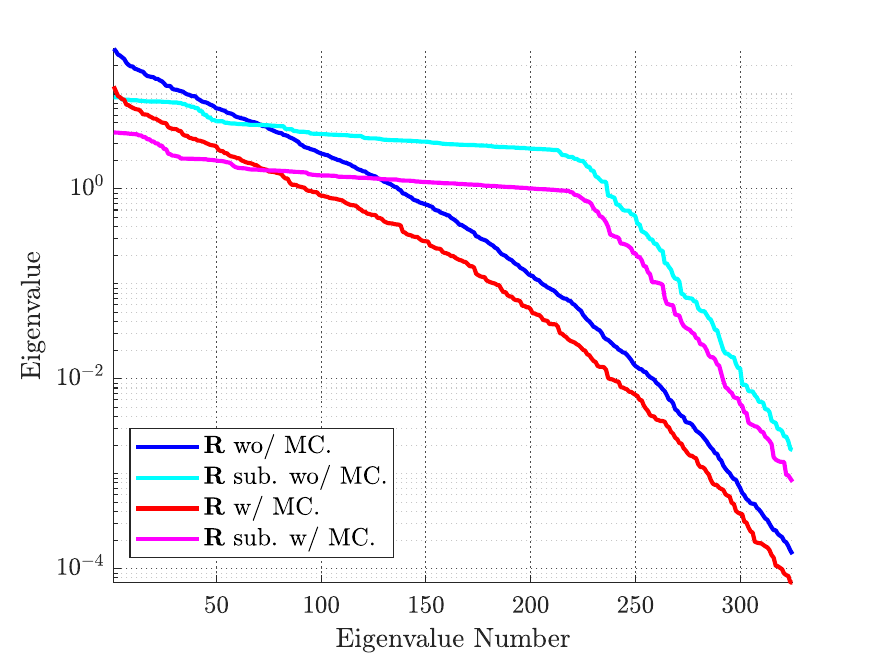}}%
 \caption{Comparison of the impact of MC on the sorted eigenvalues of exact and subspace-based spatial correlation matrices for various inter-element spacings, with a fixed \ac{ris} size of \( K_\mathrm{H}=K_\mathrm{V}=32\).}
 \label{fig:mainresults}
 \vspace{-5mm}
\end{figure*}

The channel estimation of \ac{ris}-aided links involves estimating the cascaded channel $\cv$ from the measurements in \eqref{meq}.

\subsection{LS Estimator}
\label{sec:ls_est}

The \ac{ls} approach is a widely used method for channel estimation \cite{Mishra2019Channel, Jensen2020Optimal}. To uniquely estimate \( \cv \) using a transmitted pilot symbol with power \( p_t \), the \ac{ls} estimator is given by
\begin{equation}
\hat{\cv}_{\mathrm{LS}} = \frac{1}{\sqrt{p_t}}(\Qm^\Hpow \Qm)^{-1}\Qm^\Hpow\yv,
\end{equation}
where the error covariance matrix is expressed as \cite[eq. (11)]{Pan2022Overview}
\begin{equation}
\Rm_{e_\lsidx}=  \frac{1}{\gamma N}(\Phim^\Tpow \Phim^\Strpow)^{-1} \otimes \mbf{I}_{MN},
\end{equation}
and we define the \ac{snr} as \(\gamma = p_t/\sigma^2\).

\subsection{MMSE Estimator}
\label{sec:mmse_est}

The \ac{mmse} estimator minimizes the mean square error and relies on the second-order statistics of the channel and noise. The \ac{mmse} estimate of \( \cv \) is given by \cite[eq. (12.26)]{Kay1993Fundamentals}
\begin{equation}
\hat{\cv}_{\mathrm{MMSE}} = \sqrt{p_t}\Rm_{\cpow\cpow}\Qm^\Hpow(p_t\Qm\Rm_{\cpow\cpow}\Qm^\Hpow + \sigma^2 \mbf{I}_{M T_\ppow})^{-1}\yv,
\end{equation}
where the covariance matrix of the cascaded channel is \cite{Pan2022Overview} 
\begin{equation}
\Rm_{\cpow\cpow} = (\Rm_{\Ht\Rt} \odot \Rm_{\Ft\Rt}) \otimes \Rm_{\Ht\Ut} \otimes \Rm_{\Ft\Bt}.
\end{equation}
The \ac{mmse} error covariance matrix is \cite[eq. (12.29)]{Kay1993Fundamentals}
\begin{equation}
\Rm_{e_{\mathrm{MMSE}}} = \left( \Rm_{\cpow\cpow}^{-1} + \gamma \Qm^\Hpow \Qm \right)^{-1}= \left( \Rm_{\cpow\cpow}^{-1} + \gamma N (\Phim^\Tpow \Phim^\Strpow) \otimes \mbf{I}_{MN} \right)^{-1}.
\end{equation}

\subsection{RS-LS Estimator}
\label{sec:rsls_est}

The \ac{RS-LS} estimator, proposed in \cite{Demir2022Exploiting}, mitigates pilot overhead in \ac{ris}-aided systems while improving channel estimation accuracy. Instead of estimating the exact correlation matrix, it considers a subspace spanned by another spatial
correlation matrix that represents the union of the span of all plausible correlation matrices. This approach reduces computational complexity and pilot overhead, particularly in \ac{mimo} systems, where traditional estimators require additional training. 

The eigen decomposition of the channel spatial correlation matrix is $\Rm_{\cpow\cpow} \!=\! \Um \Dm \Um^\Hpow$, where \( \Um \) contains the eigenvectors and \( \Dm \) is the diagonal matrix of eigenvalues. The \ac{RS-LS} estimate of the channel vector \( \cv \) is obtained by projecting the received signal onto a reduced subspace spanned
by \( \Um_1 \), which consists of the orthonormal eigenvectors corresponding to the nonzero eigenvalues of \( \Rm_{\cpow\cpow} \) and is structured as the Kronecker product of the BS, RIS, and UE subspaces,
\begin{equation}
\Um_1 = \Um_{\Rt\It\St,1} \otimes \Um_{\Ut\Et,1} \otimes \Um_{\Bt\St,1},
\end{equation}
where \( \Um_{\Rt\It\St,1} \), \( \Um_{\Ut\Et,1} \), and \( \Um_{\Bt\St,1} \) represent the orthonormal eigenvectors associated with the spatial correlation matrices at the \ac{ris}, \ac{UE}, and \ac{BS}, respectively. The rank \( r \) of \( \Rm_{\cpow\cpow} \) is given by the product of the ranks of the \ac{ris}, \ac{UE}, and \ac{BS} correlation matrices, \(r \!=\! r_{\text{RIS}} \cdot r_{\text{UE}} \cdot r_{\text{BS}}\). These dimensions are significantly smaller than the original sizes and further decrease with higher correlation. Such dimensionality reduction significantly decreases the estimation's computational complexity. Moreover, \ac{MC} introduces additional correlation among unit cells, which affects the distribution of channel eigenvalues by concentrating energy into fewer dominant modes. Although standard \ac{RS-LS} assumes ideal, uncoupled arrays, we incorporate \ac{MC}-aware correlation matrices to ensure that the constructed subspace accurately reflects the true signal structure, thus improving channel estimation accuracy.

The \ac{RS-LS} estimate is
\begin{equation}
\hat{\cv}_{\text{RS-LS}} = \frac{1}{\sqrt{p_t}} \Um_1 \left( \Um_1^\Hpow \Qm^\Hpow \Qm \Um_1 \right)^{-1} \Um_1^\Hpow \Qm^\Hpow \mathbf{y}.
\end{equation}
We derive the general form of the RS-LS error covariance matrix for a \ac{mimo} \ac{UE} and \ac{BS} in a \ac{ris}-aided link as
\begin{equation}
\Rm_{e_{\text{RS-LS}}} \!=\! \frac{1}{\gamma N} \left( \Um_{\text{RIS},1}\left( \Um_{\text{RIS},1}^\Hpow (\Phim^\Tpow \Phim^\Strpow) \Um_{\text{RIS},1} \right)^{-1} \!\Um_{\text{RIS},1}^\Hpow \otimes \mbf{I}_{NM}\!\right),
\end{equation}
where the detailed derivation is included in Appendix~\ref{sec:appendix_a}.

\section{Numerical Results}

This section studies the impact of \ac{MC} on the eigenvalues in different scenarios and evaluates the \ac{NMSE}, \( \text{NMSE} = \text{Tr}(\Rm_e)/\text{Tr}(\Rm_{\cpow\cpow}) \), versus \ac{snr} for different channel estimators. The exact near-field correlation is computed by extending the clustered correlation in \cite{Demir2022Channel} to incorporate spherical wave characteristics, considering \( N \!=\! 10 \) random scattering clusters within the region $\varphi^{\apow\zpow} \!\in\! \left[ -\frac{\pi}{2}, \frac{\pi}{2} \right]$, $\varphi^{\epow\lpow} \!\in \! \left[ -\frac{\pi}{2}, \frac{\pi}{2} \right]$, and $d \!\in\! \left[ 10\text{m}, 20\text{m} \right]$. The subspace correlation is computed as in \cite[eq. (10)]{Demir2024Spatial}. The carrier frequency is set to \( f_c = 3 \) GHz.

Fig.~\ref{fig:mainresults} studies the impact of \ac{MC} on the eigenvalue distributions of exact and subspace spatial correlation matrices for different inter-element spacings. In Fig.~\ref{fig:eig_0.5}, where \( \delta =  \lambda/2 \), the subspace correlation matrix exhibits a broader eigenvalue distribution with energy spread across more components. As expected, the influence of \ac{MC} is minimal under this configuration since its effect is negligible. In contrast, Fig.~\ref{fig:eig_0.25} with much smaller spacing, \( \delta = \lambda/4 \), the \ac{MC} effect is more pronounced, compressing the eigenvalue range and concentrating energy in lower eigenvalues, highlighting the increased impact of \ac{MC} in smaller antenna spacings. Therefore, the increased \ac{MC} levels, using smaller element spacings, significantly reduce the available spatial DoF.

\begin{figure}
    \centering
    \includegraphics[width=0.9\linewidth]{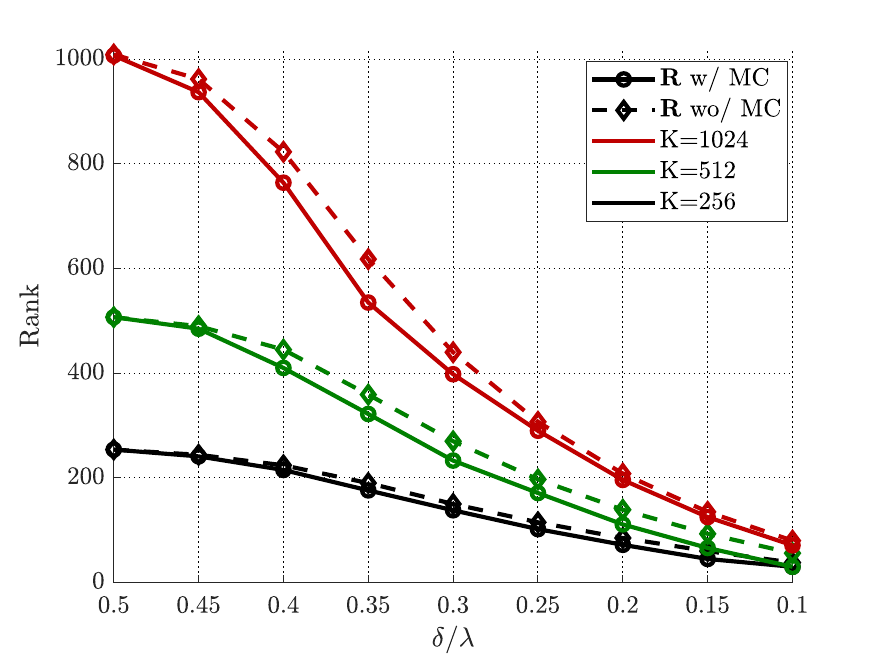}
    \caption{Impact of \ac{MC} on the rank of exact spatial correlation matrices versus inter-element spacing.}
    \label{fig:Rank_R}
    \vspace{-3mm}
\end{figure}

\begin{figure}
    \centering
    \includegraphics[width=0.9\linewidth]{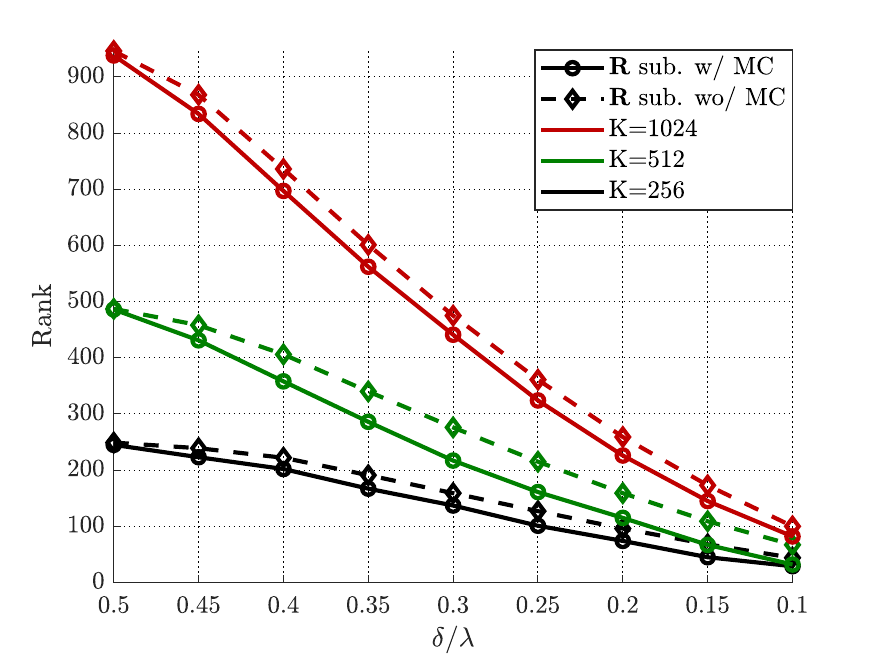}
    \caption{Impact of \ac{MC} on the rank of subspace-based spatial correlation matrices versus inter-element spacing.}
    \label{fig:Rank_sub}
    \vspace{-3mm}
\end{figure}

In Figures~\ref{fig:Rank_R} and~\ref{fig:Rank_sub}, we study the rank of the spatial correlation matrix as a function of the normalized inter-element spacing \( \delta/\lambda \) for different \ac{ris} sizes \( K \in \{256, 512, 1024\} \), with and without accounting for \ac{MC}. Specifically, Figure~\ref{fig:Rank_R} depicts the rank of the exact spatial correlation matrix, while Figure~\ref{fig:Rank_sub} presents the rank based on the subspace-based correlation. The results show that as the element spacing decreases, the rank significantly drops, indicating a reduction in the DoF due to increased spatial correlation, where this effect is more pronounced for larger \ac{ris} sizes. In particular, the presence of \ac{MC} further reduces the rank across all configurations, especially at smaller spacings, highlighting its detrimental impact on channel richness and emphasizing the importance of such accurate modeling in dense configurations.

Fig.~\ref{fig:SNR} illustrates the estimators' NMSE performance as a function of \ac{snr}. The simulation parameters are \( K = 100 \), \( M = 16 \), \( N = 4 \), \( \delta_\Bt = \frac{\lambda}{4} \), and \( \delta_\Rt = \delta_\Ut = \frac{\lambda}{8} \). The results show that the \ac{ls} estimator performs poorly, especially in systems affected by \ac{MC} effects. In contrast, the \ac{RS-LS} estimator significantly improves performance, particularly when incorporating \ac{MC} effects into the correlation matrix, even with subspace-based matrices. The additional \ac{MC} information enhances estimation accuracy, with the \ac{RS-LS} estimator that accounts for both \ac{MC} and spatial correlation outperforming its counterpart without \ac{MC} information by approximately $\unit[5]{dB}$ in NMSE.

Figure~\ref{fig:spacing_performance} shows the impact of inter-element spacing on spatial correlation, and subsequently the estimation accuracy. The simulation parameters are \( K = 128 \), \( M = 8 \), \( N = 4 \), with fixed \( p_t \). A key observation is that at smaller spacings, stronger correlation improves estimation performance. Notably, for closely spaced elements (e.g., \( \delta_\Rt = \delta_\Bt = \delta_\Ut = \frac{\lambda}{20} \)), the \ac{RS-LS} estimator achieves significantly better accuracy, outperforming the \ac{ls} estimator by several dBs. This highlights the benefit of properly exploiting \ac{MC} at the \ac{ris} and \ac{BS}. However, as spacing increases, the performance of both estimators converges, indicating a reduced impact of spatial correlation.

\begin{figure}
   \centering
\includegraphics[width=0.95\linewidth]{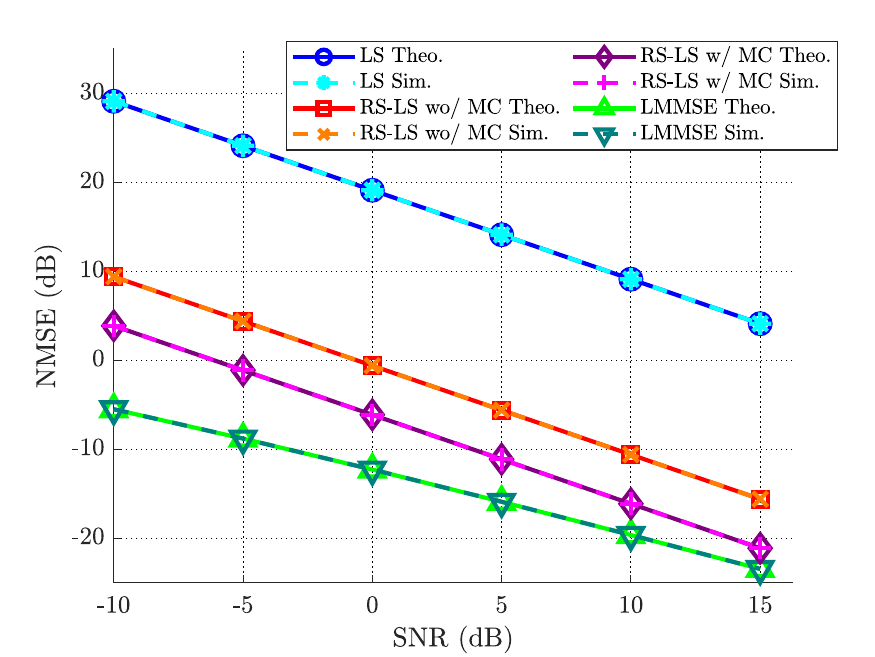}
    \caption{NMSE performance versus SNR for different channel estimators.} 
    \label{fig:SNR}
    \vspace{-3mm}
\end{figure}

\begin{figure}
   \centering
   \includegraphics[width=0.95\linewidth]{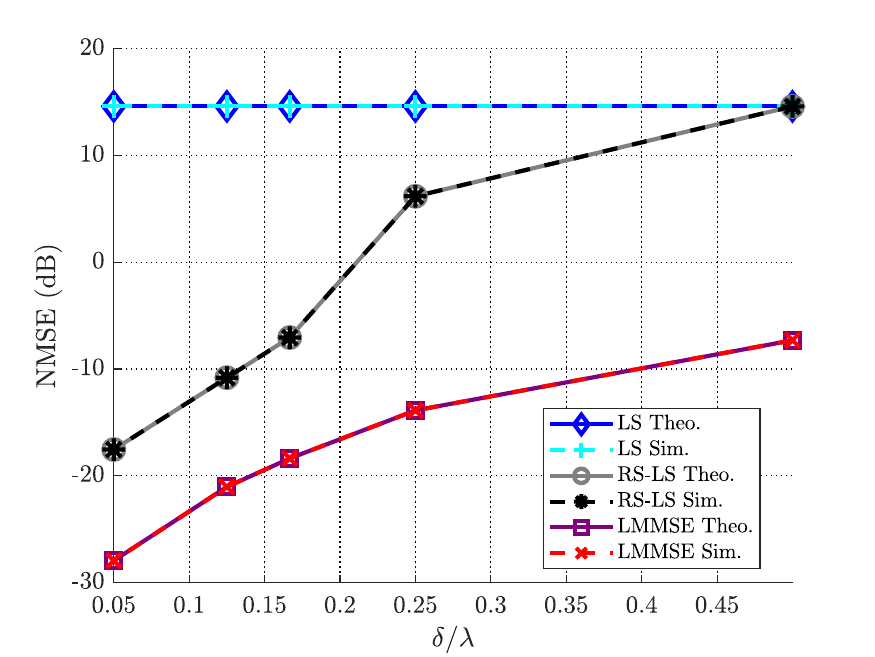}
    \caption{Comparison of estimation accuracy for varying inter-element spacing.}
    \label{fig:spacing_performance}
    \vspace{-3mm}
\end{figure}

\section{Conclusions}\label{sec:conc}

This study presents a comprehensive near-field channel modeling and estimation framework that accurately incorporates \ac{MC} effects at the \ac{ris}, \ac{BS}, and \ac{UE}. We develop and analyze the \ac{RS-LS} estimator for \ac{ris}-aided \ac{mimo} systems, leveraging \ac{MC} information to enhance estimation accuracy. In addition, we derive a general formula for the error covariance matrix of the \ac{RS-LS} estimator in \ac{mimo} \ac{ris} systems. Our results demonstrate significant performance gains with the \ac{RS-LS} estimator, particularly at smaller inter-element spacings. These findings confirm that the \ac{RS-LS} estimator effectively exploits \ac{MC} information in large-scale \ac{ris}-aided systems. 

\section{Acknowledgments}
We thank Prof. Emil Björnson for his valuable feedback.

\appendices
\counterwithin*{equation}{section}
\renewcommand\theequation{\thesection.\arabic{equation}}
\section{RS-LS error covariance matrix}
\label{sec:appendix_a}

The RS-LS error covariance matrix is~\cite[eq. (8)]{Demir2022Exploiting}
\begin{equation}
\Rm_{e_{\text{RS-LS}}} = \frac{1}{\gamma} \Um_1 \left( 
\Um_1^\Hpow \Qm^\Hpow \Qm \Um_1 
\right)^{-1} \Um_1^\Hpow,
\end{equation}
where we can write by following~\cite[eq. (11)]{Swindlehurst2022Channel},
\begin{equation}
    \Qm^\Hpow \Qm = N (\Phim^\Tpow \Phim^\Strpow) \otimes \mathbf{I}_{NM}.
\end{equation}
Then, by using $\Um_1$, the inverse term is expressed as
\begin{equation}
\begin{split}
    &\left(\Um_1^\Hpow \Qm^\Hpow \Qm \Um_1\right)^{-1} = \Big( \left( \Um_{\Rt\It\St,1} \otimes \Um_{\Ut\Et,1} \otimes \Um_{\Bt\St,1} \right)^\Hpow \\
    &\times\left(N (\Phim^\Tpow \Phim^\Strpow) \otimes \mathbf{I}_{N} \otimes \mathbf{I}_{M}\right)  \left( \Um_{\Rt\It\St,1} \otimes \Um_{\Ut\Et,1} \otimes \Um_{\Bt\St,1} \right) \Big)^{-1} \\
    &=\!N\!\left(\!\Um_{\Rt\It\St,\!1}^\Hpow\!(\!\Phim^\Tpow \Phim^\Strpow\!) \Um_{\Rt\It\St,\!1}\!\right)\!\otimes\!\left(\!\Um_{\Ut\Et,1}^\Hpow\!\mathbf{I}_{N}\!\Um_{\Ut\Et,1}\!\right) \otimes \left(\!\Um_{\Bt\St,1}^\Hpow\!\mathbf{I}_{M} \Um_{\Bt\St,\!1}\!\right)^{-1} \\
    &= \frac{1}{N} \left( \Um_{\Rt\It\St,1}^\Hpow (\Phim^\Tpow \Phim^\Strpow) \Um_{\Rt\It\St,1} \right)^{-1} \otimes \mathbf{I}_{NM}.
\end{split}
\end{equation}
To complete the expression, we evaluate
\begin{equation}
\begin{split}
    & \Um_1 \left( \Um_1^\Hpow \Qm^\Hpow \Qm \Um_1 \right)^{-1} \Um_1^\Hpow  = \left( \Um_{\Rt\It\St,1} \otimes \Um_{\Ut\Et,1} \otimes \Um_{\Bt\St,1} \right) \\
    & \times \left( \frac{1}{N} \left( \Um_{\Rt\It\St,1}^\Hpow (\Phim^\Tpow \Phim^\Strpow) \Um_{\Rt\It\St,1} \right)^{-1} \otimes \mathbf{I}_{NM} \right) \\
    & \times \left( \Um_{\Rt\It\St,1}^\Hpow \otimes \Um_{\Ut\Et,1}^\Hpow \otimes \Um_{\Bt\St,1}^\Hpow \right),
\end{split}
\end{equation}
where applying the Kronecker properties results in
\begin{equation}
    \frac{1}{N} \left( \Um_{\Rt\It\St,1} \left( \Um_{\Rt\It\St,1}^\Hpow (\Phim^\Tpow \Phim^\Strpow) \Um_{\Rt\It\St,1} \right)^{-1} \Um_{\Rt\It\St,1}^\Hpow \right) \otimes \mathbf{I}_{NM}.
\end{equation}
The RS-LS error covariance matrix is finally expressed as
\begin{equation}
    \Rm_{e_{\text{RS-LS}}} \!=\! \frac{1}{\gamma N} \left( \Um_{\text{RIS},1} \left( \Um_{\text{RIS},1}^\Hpow (\Phim^\Tpow \Phim^\Strpow) \Um_{\text{RIS},1} \right)^{-1} \Um_{\text{RIS},1}^\Hpow  \otimes \mathbf{I}_{NM}\right).
\end{equation}

\bibliography{IEEEabrv,ref}
\bibliographystyle{IEEEtran}
\end{document}

%% file: macros.tex
\newcommand{\red}[1]{{\color{red}{#1}}} 
\newcommand{\blue}[1]{{\color{blue}{#1}}} 
\newcommand{\green}[1]{{\color{green}{#1}}} 
\newcommand{\yellow}[1]{{\color{yellow}{#1}}} 
\newcommand{\orange}[1]{{\color{orange}{#1}}} 

\newcommand{\nth}[1]{{#1}{\text{th}}}
\newcommand{\mbf}[1]{\mathbf{#1}}

\newcommand{\Hpow}{{\sf H}}
\newcommand{\Vpow}{{\sf V}}
\newcommand{\Tpow}{{\sf T}}
\newcommand{\Invpow}{{\sf -1}}
\newcommand{\Strpow}{{\sf *}}

\newcommand{\PseuInvpow}{\mathrm{\dagger}}
\newcommand{\apow}{\mathrm{a}}
\newcommand{\zpow}{\mathrm{z}}
\newcommand{\lpow}{\mathrm{l}}
\newcommand{\npow}{\mathrm{n}}
\newcommand{\Npow}{\mathrm{N}}
\newcommand{\vpow}{\mathrm{v}}
\newcommand{\cpow}{\mathrm{c}}
\newcommand{\ipow}{\mathrm{i}}
\newcommand{\jpow}{\mathrm{j}}
\newcommand{\bpow}{\mathrm{b}}
\newcommand{\rpow}{\mathrm{r}}
\newcommand{\upow}{\mathrm{u}}
\newcommand{\Upow}{\mathrm{U}}
\newcommand{\Bpow}{\mathrm{B}}
\newcommand{\Rpow}{\mathrm{R}}
\newcommand{\tpow}{\mathrm{t}}
\newcommand{\hpow}{\mathrm{h}}
\newcommand{\epow}{\mathrm{e}}
\newcommand{\dpow}{\mathrm{d}}
\newcommand{\Dpow}{\mathrm{D}}
\newcommand{\ppow}{\mathrm{p}}
\newcommand{\kpow}{\mathrm{k}}
\newcommand{\Mpow}{\mathrm{M}}
\newcommand{\mpow}{\mathrm{m}}
\newcommand{\Kpow}{\mathrm{K}}
\newcommand{\Fpow}{\mathrm{F}}
\newcommand{\qpow}{\mathrm{q}}
\newcommand{\abs}[1]{\left|{#1}\right|}
\newcommand{\norm}[1]{\left\|{#1}\right\|}
\newcommand{\vect}[1]{\mathrm{vec}\left(#1\right)}
\newcommand{\supp}[1]{\mathrm{supp}\left(#1\right)}
\newcommand{\trc}[1]{\mathrm{tr}\left(#1\right)}
\newcommand{\diagopr}[1]{\mathrm{diag}\left(#1\right)}
\newcommand{\blkdiagopr}[1]{\mathrm{blkdiag}\left(#1\right)}

\newcommand{\atantwo}{\mathrm{arctan2}}

\newcommand{\lmmseidx}{\mathrm{LMMSE}}
\newcommand{\lsidx}{\mathrm{LS}}
\newcommand{\txidx}{\mathrm{T}}
\newcommand{\rxidx}{\mathrm{R}}
\newcommand{\strmidx}{\mathrm{S}}
\newcommand{\froidx}{\mathrm{F}}
\newcommand{\sysidx}{\mathrm{sys}}
\newcommand{\RFidx}{\mathrm{RF}}
\newcommand{\BBidx}{\mathrm{BB}}
\newcommand{\quantidx}{\mathrm{quant}}
\newcommand{\effidx}{\mathrm{eff}}
\newcommand{\totsupsc}{\mathrm{tot}}
\newcommand{\firstelement}{\mathrm{st}}

\newcommand{\pilotidx}{\mathrm{p}}
\newcommand{\Beamsupsc}{\mathrm{beam}}
\newcommand{\CP}{\mathrm{CP}}
\newcommand{\cent}{\mathrm{cen}}

\newcommand{\coh}{\mathrm{coh}}
\newcommand{\train}{\mathrm{tr}}
\newcommand{\ovs}{\mathrm{ovs}}
\newcommand{\DicRed}{\mathrm{RD}}
\newcommand{\LessCol}{\mathrm{lc}}

\newcommand{\clu}{\mathrm{clu}}
\newcommand{\ray}{\mathrm{ray}}
\newcommand{\LoS}{\mathrm{los}}
\newcommand{\NLoS}{\mathrm{nlos}}
\newcommand{\subb}{\mathrm{sub}}
\newcommand{\GMM}{\mathrm{GMM}}
\newcommand{\svv}{\mathrm{sv}}

\newcommand{\samp}{\mathrm{s}}
\newcommand{\ula}{\mathrm{ULA}}
\newcommand{\upa}{\mathrm{UPA}}
\newcommand{\bmsp}{\mathrm{bmsp}}

\newcommand{\NFsupsc}{\mathrm{NF}}
\newcommand{\FFsupsc}{\mathrm{FF}}
\newcommand{\SWMsupsc}{\mathrm{SWM}}
\newcommand{\HSPMsupsc}{\mathrm{HSPWM}}
\newcommand{\PWMsupsc}{\mathrm{PWM}}

\newcommand{\umarridx}{\mathrm{UMA}}
\newcommand{\saidx}{\mathrm{SA}}
\newcommand{\aeidx}{\mathrm{AE}}

\newcommand{\rotsupsc}{\mathrm{rot}}
\newcommand{\trialsupsc}{\mathrm{trl}}
\newcommand{\maxsupsc}{\mathrm{max}}
\newcommand{\Polsupsc}{\mathrm{pol}}
\newcommand{\DFTsupsc}{\mathrm{dft}}

\newcommand{\allidx}{\mathrm{all}}
\newcommand{\distsupsc}{\mathrm{dist}}
\newcommand{\offsupsc}{\mathrm{offline}}
\newcommand{\onsupsc}{\mathrm{online}}
\newcommand{\estidx}{\mathrm{est}}
\setlength\unitlength{1mm}

\newcommand{\insertfig}[3]{
\begin{figure}[htbp]\begin{center}\begin{picture}(120,90)
\put(0,-5){\includegraphics[width=12cm,height=9cm,clip=]{#1.eps}}\end{picture}\end{center}
\caption{#2}\label{#3}\end{figure}}

\newcommand{\insertxfig}[4]{
\begin{figure}[htbp]
\begin{center}
\leavevmode \centerline{\resizebox{#4\textwidth}{!}{\input
#1.pstex_t}}
\caption{#2} \label{#3}
\end{center}
\end{figure}}

\long\def\comment#1{}



\newfont{\bbb}{msbm10 scaled 700}
\newcommand{\CCC}{\mbox{\bbb C}}

\newfont{\bb}{msbm10 scaled 1100}
\newcommand{\CC}{\mbox{\bb C}}
\newcommand{\PP}{\mbox{\bb P}}
\newcommand{\RR}{\mbox{\bb R}}
\newcommand{\QQ}{\mbox{\bb Q}}
\newcommand{\ZZ}{\mbox{\bb Z}}
\newcommand{\FF}{\mbox{\bb F}}
\newcommand{\GG}{\mbox{\bb G}}
\newcommand{\EE}{\mbox{\bb E}}
\newcommand{\NN}{\mbox{\bb N}}
\newcommand{\KK}{\mbox{\bb K}}


\newcommand{\av}{{\bf a}}
\newcommand{\bv}{{\bf b}}
\newcommand{\cv}{{\bf c}}
\newcommand{\dv}{{\bf d}}
\newcommand{\ev}{{\bf e}}
\newcommand{\fv}{{\bf f}}
\newcommand{\gv}{{\bf g}}
\newcommand{\hv}{{\bf h}}
\newcommand{\iv}{{\bf i}}
\newcommand{\jv}{{\bf j}}
\newcommand{\kv}{{\bf k}}
\newcommand{\lv}{{\bf l}}
\newcommand{\mv}{{\bf m}}
\newcommand{\nv}{{\bf n}}
\newcommand{\ov}{{\bf o}}
\newcommand{\pv}{{\bf p}}
\newcommand{\qv}{{\bf q}}
\newcommand{\rv}{{\bf r}}
\newcommand{\sv}{{\bf s}}
\newcommand{\tv}{{\bf t}}
\newcommand{\uv}{{\bf u}}
\newcommand{\wv}{{\bf w}}
\newcommand{\xv}{{\bf x}}
\newcommand{\yv}{{\bf y}}
\newcommand{\zv}{{\bf z}}
\newcommand{\zerov}{{\bf 0}}
\newcommand{\onev}{{\bf 1}}

\def\u{{\bf u}}


\newcommand{\Am}{{\bf A}}
\newcommand{\Bm}{{\bf B}}
\newcommand{\Cm}{{\bf C}}
\newcommand{\Dm}{{\bf D}}
\newcommand{\Em}{{\bf E}}
\newcommand{\Fm}{{\bf F}}
\newcommand{\Gm}{{\bf G}}
\newcommand{\Hm}{{\bf H}}
\newcommand{\Id}{{\bf I}}
\newcommand{\Jm}{{\bf J}}
\newcommand{\Km}{{\bf K}}
\newcommand{\Lm}{{\bf L}}
\newcommand{\Mm}{{\bf M}}
\newcommand{\Nm}{{\bf N}}
\newcommand{\Om}{{\bf O}}
\newcommand{\Pm}{{\bf P}}
\newcommand{\Qm}{{\bf Q}}
\newcommand{\Rm}{{\bf R}}
\newcommand{\Sm}{{\bf S}}
\newcommand{\Tm}{{\bf T}}
\newcommand{\Um}{{\bf U}}
\newcommand{\Wm}{{\bf W}}
\newcommand{\Vm}{{\bf V}}
\newcommand{\Xm}{{\bf X}}
\newcommand{\Ym}{{\bf Y}}
\newcommand{\Zm}{{\bf Z}}
\newcommand{\Onem}{{\bf 1}}
\newcommand{\Zerom}{{\bf 0}}


\newcommand{\Bc}{{\cal B}}
\newcommand{\Cc}{{\cal C}}
\newcommand{\Dc}{{\cal D}}
\newcommand{\Ec}{{\cal E}}
\newcommand{\Fc}{{\cal F}}
\newcommand{\Gc}{{\cal G}}
\newcommand{\Hc}{{\cal H}}
\newcommand{\Ic}{{\cal I}}
\newcommand{\Jc}{{\cal J}}
\newcommand{\Kc}{{\cal K}}
\newcommand{\Lc}{{\cal L}}
\newcommand{\Mc}{{\cal M}}
\newcommand{\Nc}{{\cal N}}
\newcommand{\Oc}{{\cal O}}
\newcommand{\Pc}{{\cal P}}
\newcommand{\Qc}{{\cal Q}}
\newcommand{\Rc}{{\cal R}}
\newcommand{\Sc}{{\cal S}}
\newcommand{\Tc}{{\cal T}}
\newcommand{\Uc}{{\cal U}}
\newcommand{\Wc}{{\cal W}}
\newcommand{\Vc}{{\cal V}}
\newcommand{\Xc}{{\cal X}}
\newcommand{\Yc}{{\cal Y}}
\newcommand{\Zc}{{\cal Z}}


\newcommand{\alphav}{\hbox{\boldmath$\alpha$}}
\newcommand{\betav}{\hbox{\boldmath$\beta$}}
\newcommand{\gammav}{\hbox{\boldmath$\gamma$}}
\newcommand{\deltav}{\hbox{\boldmath$\delta$}}
\newcommand{\etav}{\hbox{\boldmath$\eta$}}
\newcommand{\lambdav}{\hbox{\boldmath$\lambda$}}
\newcommand{\epsilonv}{\hbox{\boldmath$\epsilon$}}
\newcommand{\nuv}{\hbox{\boldmath$\nu$}}
\newcommand{\muv}{\hbox{\boldmath$\mu$}}
\newcommand{\zetav}{\hbox{\boldmath$\zeta$}}
\newcommand{\phiv}{\hbox{\boldmath$\phi$}}
\newcommand{\psiv}{\hbox{\boldmath$\psi$}}
\newcommand{\thetav}{\hbox{\boldmath$\theta$}}
\newcommand{\tauv}{\hbox{\boldmath$\tau$}}
\newcommand{\omegav}{\hbox{\boldmath$\omega$}}
\newcommand{\xiv}{\hbox{\boldmath$\xi$}}
\newcommand{\sigmav}{\hbox{\boldmath$\sigma$}}
\newcommand{\piv}{\hbox{\boldmath$\pi$}}
\newcommand{\rhov}{\hbox{\boldmath$\rho$}}
\newcommand{\vtv}{\hbox{\boldmath$\vartheta$}}

\newcommand{\Gammam}{\hbox{\boldmath$\Gamma$}}
\newcommand{\Lambdam}{\hbox{\boldmath$\Lambda$}}
\newcommand{\Deltam}{\hbox{\boldmath$\Delta$}}
\newcommand{\Sigmam}{\hbox{\boldmath$\Sigma$}}
\newcommand{\Phim}{\hbox{\boldmath$\Phi$}}
\newcommand{\Pim}{\hbox{\boldmath$\Pi$}}
\newcommand{\Psim}{\hbox{\boldmath$\Psi$}}
\newcommand{\psim}{\hbox{\boldmath$\psi$}}
\newcommand{\chim}{\hbox{\boldmath$\chi$}}
\newcommand{\omegam}{\hbox{\boldmath$\omega$}}
\newcommand{\vphim}{\hbox{\boldmath$\varphi$}}
\newcommand{\Thetam}{\hbox{\boldmath$\Theta$}}
\newcommand{\Omegam}{\hbox{\boldmath$\Omega$}}
\newcommand{\Xim}{\hbox{\boldmath$\Xi$}}


\newcommand{\sinc}{{\hbox{sinc}}}
\newcommand{\diag}{{\hbox{diag}}}
\renewcommand{\det}{{\hbox{det}}}
\newcommand{\trace}{{\hbox{tr}}}
\newcommand{\sign}{{\hbox{sign}}}
\renewcommand{\arg}{{\hbox{arg}}}
\newcommand{\var}{{\hbox{var}}}
\newcommand{\cov}{{\hbox{cov}}}
\newcommand{\SINR}{{\sf sinr}}
\newcommand{\SNR}{{\sf snr}}
\newcommand{\Ei}{{\rm E}_{\rm i}}
\newcommand{\eqdef}{\stackrel{\Delta}{=}}
\newcommand{\defines}{{\,\,\stackrel{\scriptscriptstyle \bigtriangleup}{=}\,\,}}
\newcommand{\<}{\left\langle}
\renewcommand{\>}{\right\rangle}
\newcommand{\herm}{{\sf H}}
\newcommand{\trasp}{{\sf T}}
\renewcommand{\vec}{{\rm vec}}
\newcommand{\calL}{\mbox{${\mathcal L}$}}
\newcommand{\calO}{\mbox{${\mathcal O}$}}

\newcommand{\Afd}{\mbox{$\boldsymbol{\mathcal{A}}$}}
\newcommand{\Bfd}{\mbox{$\boldsymbol{\mathcal{B}}$}}
\newcommand{\Cfd}{\mbox{$\boldsymbol{\mathcal{C}}$}}
\newcommand{\Dfd}{\mbox{$\boldsymbol{\mathcal{D}}$}}
\newcommand{\Efd}{\mbox{$\boldsymbol{\mathcal{E}}$}}
\newcommand{\Ffd}{\mbox{$\boldsymbol{\mathcal{F}}$}}
\newcommand{\Gfd}{\mbox{$\boldsymbol{\mathcal{G}}$}}
\newcommand{\Hfd}{\mbox{$\boldsymbol{\mathcal{H}}$}}
\newcommand{\Ifd}{\mbox{$\boldsymbol{\mathcal{I}}$}}
\newcommand{\Jfd}{\mbox{$\boldsymbol{\mathcal{J}}$}}
\newcommand{\Kfd}{\mbox{$\boldsymbol{\mathcal{K}}$}}
\newcommand{\Lfd}{\mbox{$\boldsymbol{\mathcal{L}}$}}
\newcommand{\Mfd}{\mbox{$\boldsymbol{\mathcal{M}}$}}
\newcommand{\Nfd}{\mbox{$\boldsymbol{\mathcal{N}}$}}
\newcommand{\Ofd}{\mbox{$\boldsymbol{\mathcal{O}}$}}
\newcommand{\Pfd}{\mbox{$\boldsymbol{\mathcal{P}}$}}
\newcommand{\Qfd}{\mbox{$\boldsymbol{\mathcal{Q}}$}}
\newcommand{\Rfd}{\mbox{$\boldsymbol{\mathcal{R}}$}}
\newcommand{\Sfd}{\mbox{$\boldsymbol{\mathcal{S}}$}}
\newcommand{\Tfd}{\mbox{$\boldsymbol{\mathcal{T}}$}}
\newcommand{\Ufd}{\mbox{$\boldsymbol{\mathcal{U}}$}}
\newcommand{\Vfd}{\mbox{$\boldsymbol{\mathcal{V}}$}}
\newcommand{\Wfd}{\mbox{$\boldsymbol{\mathcal{W}}$}}
\newcommand{\Xfd}{\mbox{$\boldsymbol{\mathcal{X}}$}}
\newcommand{\Yfd}{\mbox{$\boldsymbol{\mathcal{Y}}$}}
\newcommand{\Zfd}{\mbox{$\boldsymbol{\mathcal{Z}}$}}

\newcommand{\At}{{\rm A}}
\newcommand{\Bt}{{\rm B}}
\newcommand{\Ct}{{\rm C}}
\newcommand{\Dt}{{\rm D}}
\newcommand{\Et}{{\rm E}}
\newcommand{\Ft}{{\rm F}}
\newcommand{\Gt}{{\rm G}}
\newcommand{\Ht}{{\rm H}}
\newcommand{\It}{{\rm I}}
\newcommand{\Jt}{{\rm J}}
\newcommand{\Kt}{{\rm K}}
\newcommand{\Lt}{{\rm L}}
\newcommand{\Mt}{{\rm M}}
\newcommand{\Nt}{{\rm N}}
\newcommand{\Ot}{{\rm O}}
\newcommand{\Pt}{{\rm P}}
\newcommand{\Qt}{{\rm Q}}
\newcommand{\Rt}{{\rm R}}
\newcommand{\St}{{\rm S}}
\newcommand{\Tt}{{\rm T}}
\newcommand{\Ut}{{\rm U}}
\newcommand{\Wt}{{\rm W}}
\newcommand{\Vt}{{\rm V}}
\newcommand{\Xt}{{\rm X}}
\newcommand{\Yt}{{\rm Y}}
\newcommand{\Zt}{{\rm Z}}

\newcommand{\varphiv}{\hbox{\boldmath$\varphi$}}


\def\bbA{{\mathbb{A}}}
\def\bbB{{\mathbb{B}}}
\def\bbC{{\mathbb{C}}}
\def\bbD{{\mathbb{D}}}
\def\bbE{{\mathbb{E}}}
\def\bbF{{\mathbb{F}}}
\def\bbG{{\mathbb{G}}}
\def\bbH{{\mathbb{H}}}
\def\bbI{{\mathbb{I}}}
\def\bbJ{{\mathbb{J}}}
\def\bbK{{\mathbb{K}}}
\def\bbL{{\mathbb{L}}}
\def\bbM{{\mathbb{M}}}
\def\bbN{{\mathbb{N}}}
\def\bbO{{\mathbb{O}}}
\def\bbP{{\mathbb{P}}}
\def\bbQ{{\mathbb{Q}}}
\def\bbR{{\mathbb{R}}}
\def\bbS{{\mathbb{S}}}
\def\bbT{{\mathbb{T}}}
\def\bbU{{\mathbb{U}}}
\def\bbV{{\mathbb{V}}}
\def\bbW{{\mathbb{W}}}
\def\bbX{{\mathbb{X}}}
\def\bbY{{\mathbb{Y}}}
\def\bbZ{{\mathbb{Z}}}

\def\cA{\mathcal{A}}
\def\cB{\mathcal{B}}
\def\cC{\mathcal{C}}
\def\cD{\mathcal{D}}
\def\cE{\mathcal{E}}
\def\cF{\mathcal{F}}
\def\cG{\mathcal{G}}
\def\cH{\mathcal{H}}
\def\cI{\mathcal{I}}
\def\cJ{\mathcal{J}}
\def\cK{\mathcal{K}}
\def\cL{\mathcal{L}}
\def\cM{\mathcal{M}}
\def\cN{\mathcal{N}}
\def\cO{\mathcal{O}}
\def\cP{\mathcal{P}}
\def\cQ{\mathcal{Q}}
\def\cR{\mathcal{R}}
\def\cS{\mathcal{S}}
\def\cT{\mathcal{T}}
\def\cU{\mathcal{U}}
\def\cV{\mathcal{V}}
\def\cW{\mathcal{W}}
\def\cX{\mathcal{X}}
\def\cY{\mathcal{Y}}
\def\cZ{\mathcal{Z}}

\def\bba{{\mathbb{a}}}
\def\bbb{{\mathbb{b}}}
\def\bbc{{\mathbb{c}}}
\def\bbd{{\mathbb{d}}}
\def\bbee{{\mathbb{e}}}
\def\bbff{{\mathbb{f}}}
\def\bbg{{\mathbb{g}}}
\def\bbh{{\mathbb{h}}}
\def\bbi{{\mathbb{i}}}
\def\bbj{{\mathbb{j}}}
\def\bbk{{\mathbb{k}}}
\def\bbl{{\mathbb{l}}}
\def\bbm{{\mathbb{m}}}
\def\bbn{{\mathbb{n}}}
\def\bbo{{\mathbb{o}}}
\def\bbp{{\mathbb{p}}}
\def\bbq{{\mathbb{q}}}
\def\bbr{{\mathbb{r}}}
\def\bbs{{\mathbb{s}}}
\def\bbt{{\mathbb{t}}}
\def\bbu{{\mathbb{u}}}
\def\bbv{{\mathbb{v}}}
\def\bbw{{\mathbb{w}}}
\def\bbx{{\mathbb{x}}}
\def\bby{{\mathbb{y}}}
\def\bbz{{\mathbb{z}}}
\def\bbzero{{\mathbb{0}}}
\def\bbone{{\mathbb{1}}}